\documentclass[%
reprint, 
superscriptaddress, 
amsmath,
amssymb,
aps,
prx,
]{revtex4-2}

\usepackage{graphicx} 
\usepackage{dcolumn} 
\usepackage{bm} 
\usepackage[pdftex,breaklinks,colorlinks=true,citecolor=blue,linkcolor=blue,urlcolor=blue]{hyperref}
\usepackage{ulem} 
\usepackage{tabularx} 
\usepackage{braket} 
\usepackage{xcolor} 
\usepackage{textcomp} 
\usepackage{amsmath} 
\usepackage{multirow}
\usepackage{comment}
\usepackage{ragged2e} 
\usepackage{natbib}

\definecolor{darkgreen}{RGB}{0, 160, 0}
\definecolor{lavender}{RGB}{160, 160, 240}

\begin{document}

\title{Site-selective preparation and multi-state readout of molecules in optical tweezers}
\author{Lewis~R.~B.~Picard} 
\thanks{A.J.P., G.E.P. and L.R.B.P contributed equally to this work.}
\affiliation{Department of Physics, Harvard University, Cambridge, Massachusetts 02138, USA}
\affiliation{Department of Chemistry and Chemical Biology, Harvard University, Cambridge, Massachusetts 02138, USA}
\affiliation{Harvard-MIT Center for Ultracold Atoms, Cambridge, Massachusetts 02138, USA}

\author{Gabriel~E.~Patenotte} 
\thanks{A.J.P., G.E.P. and L.R.B.P contributed equally to this work.}
\affiliation{Department of Physics, Harvard University, Cambridge, Massachusetts 02138, USA}
\affiliation{Department of Chemistry and Chemical Biology, Harvard University, Cambridge, Massachusetts 02138, USA}
\affiliation{Harvard-MIT Center for Ultracold Atoms, Cambridge, Massachusetts 02138, USA}

\author{Annie~J.~Park}
\thanks{A.J.P., G.E.P. and L.R.B.P contributed equally to this work.}
\affiliation{Department of Chemistry and Chemical Biology, Harvard University, Cambridge, Massachusetts 02138, USA}
\affiliation{Department of Physics, Harvard University, Cambridge, Massachusetts 02138, USA}
\affiliation{Harvard-MIT Center for Ultracold Atoms, Cambridge, Massachusetts 02138, USA}

\author{Samuel~F.~Gebretsadkan} 
\affiliation{Department of Physics, Harvard University, Cambridge, Massachusetts 02138, USA}
\affiliation{Department of Chemistry and Chemical Biology, Harvard University, Cambridge, Massachusetts 02138, USA}
\affiliation{Harvard-MIT Center for Ultracold Atoms, Cambridge, Massachusetts 02138, USA}

\author{Kang-Kuen~Ni} 
\email{ni@chemistry.harvard.edu}
\affiliation{Department of Chemistry and Chemical Biology, Harvard University, Cambridge, Massachusetts 02138, USA}
\affiliation{Department of Physics, Harvard University, Cambridge, Massachusetts 02138, USA}
\affiliation{Harvard-MIT Center for Ultracold Atoms, Cambridge, Massachusetts 02138, USA}

\begin{abstract}

Polar molecules are a quantum resource with rich internal structure that can be coherently controlled. The structure, however, also makes the state preparation and measurement (SPAM) of molecules challenging. We advance the SPAM of individual molecules assembled from constituent atoms trapped in optical tweezer arrays. Sites without NaCs molecules are eliminated using high-fidelity Cs atom detection, increasing the peak molecule filling fraction of the array threefold. We site-selectively initialize the array in a rotational qubit subspace that is insensitive to differential AC Stark shifts from the optical tweezer. Lastly, we detect multiple rotational states per experimental cycle by imaging atoms after sequential state-selective dissociations. These demonstrations extend the SPAM capabilities of molecules for quantum information, simulation, and metrology.

\end{abstract}

\maketitle

\section{Introduction}

Polar molecules feature a rich internal structure with tunable long-range dipolar interactions, making them a resource for a wide range of quantum science applications. 
Coherent control of internal states~\cite{Ospelkaus2010a,Park2017,gregory_robust_2021,lin2022,burchesky2021,park_extended_2023,gregory_second-scale_2024} is key to many recent advances. The dipole-dipole interaction has been used to realize a spin-Hamiltonian \cite{Gorshkov2011a} whose site-resolved correlations have been measured in an optical lattice \cite{christakis_probing_2023}, and to produce Bell-states in optical tweezers \cite{Holland_Cheuk_2023_DDI, bao_dipolar_2023} for dipolar quantum gates \cite{Ni2018}. Control over nuclear spins has transferred their entanglement from reactants to products in chemical reactions
\cite{Liu_Ni_2023_CoherenceReactions}. In the search for the electron electric dipole moment, internal states have been polarized to generate large lab-frame electric fields and used to evaluate systematic errors \cite{Hudson_Hinds_2011_EDM, Roussy_Cornell_2023_EDM, Andreev_ACME_2018_EDM}. 
Many applications require further advances in state preparation and measurement (SPAM), which remains an outstanding challenge due to the dense electronic and vibrational structure that complicates molecule production and detection.

Nondestructive detection using fluorescence from optical cycling transitions has become an indispensable technique in atomic physics. Real-time detection of individual atoms enables the rearrangement of occupied traps to produce a densely filled array \cite{Barredo_Browaeys_2016_Rearrangement, Endres_Lukin_2016_Rearrangement,shaw_dark-state_2023}, as well as the selective readout of their internal states \cite{Martinez-Dorantes_Meschede_2018_DipoleForceHeating,Low_Senko_2023_IonQudit}. Both capabilities are crucial for studying many-body interactions and executing large quantum circuits \cite{semeghini2021probing, ringbauer_universal_2022,scholl_erasure_2023,bluvstein_logical_2023}. For molecules, a subset~\cite{Rosa2004,Tarbutt2019} do possess optical cycling transitions, enabling advances in their laser-cooling \cite{shuman_laser_2010,Truppe2017}, trapping \cite{Anderegg2019,Wu2021,vilas_optical_2023}, and rearrangement \cite{Holland_Cheuk_2023_DDI}. Other molecules can be produced by the coherent association of their constituent atoms \cite{Kohler2006,Danzl2008,Ni2008,Lang2008,Takekoshi2014,Molony2014,Park2015,Guo2016,He2020, Voges2020,Zhang_Ni_2020_MagnetoassociationTweezer, Yu_Ni_2021_OpticalAssociationTweezer,  Ruttley_Cornish_2023_Mergoassociation}, with the advantage that cold atom temperatures are preserved in the molecule creation process. However, optical cycling transitions for fluorescence or absorption imaging are generally not available. 
In general, the population of a specific state of associated molecules is detected through the adiabatic dissociation of the constituent atoms, which are then separated and imaged.
To make use of the inherent multi-level structure of molecules, three capabilities for the molecule SPAM toolbox are highly desirable: nondestructive detection while maintaining low temperature, site-resolved state preparation, and resolved readout of multiple internal states. The first capability would allow for the production of deterministically filled arrays of molecules. The latter two would facilitate experiments using the many long-lived rotational states as a qudit system or multi-level quantum simulator \cite{homeier2023antiferromagnetic, Sawant_Cornish_2020_Qudit}. Various proposals have been put forward for molecule readout via a controlled interaction between a ground state molecule and an ancilla atom or optical cavity \cite{Wolf2016,Chou2017,jamadagni_quantum_2019,guan_nondestructive_2020, Wes2020,wang_enriching_2022, Zhang_Tarbutt_2022_HybridMolRydb}. However, these methods require the introduction of a new, coherently controlled interaction between the molecule and the detection ancilla, which requires much more experimental overhead to implement.

\begin{figure}[t]
\centering
\includegraphics[width = \columnwidth]{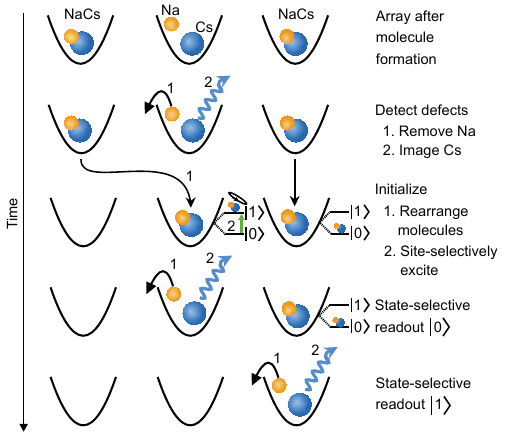}
\caption{Schematic of the SPAM techniques presented for molecular systems that enable rearrangement, site-selective state preparation, and sequential multi-state detection. Cs atoms that fail to associate with Na atoms into a NaCs molecule are detected, which is used to rearrange molecules in the rotational ground state $\ket{0}$ to a densely filled region. Site selective excitation prepares desired molecules in the excited rotational state $\ket{1}$. For multi-state detection, molecules are state-selectively dissociated and their constituent Cs atoms are imaged.}
\label{fig:SPAM_schematic_overview}
\end{figure}

Here, we demonstrate an immediately-applicable method for the removal of defects in a molecular array and sequential state-selective detection.
Our proposed scheme uses fluorescence imaging of the constituent atoms of NaCs, as shown in Fig. \ref{fig:SPAM_schematic_overview}. Unassociated Cs atoms are detected immediately following molecule formation to infer which sites contain molecules. We rearrange those sites and increase the local filling probability at the edge of the array threefold. We also use the atom signal for the detection of the two molecular states in the ground and first excited rotational levels that form our computational basis. In a single experimental cycle, molecules are sequentially dissociated from each state and the resulting Cs atoms are detected. Lastly, we demonstrate the ability to initialize an array of molecules in an arbitrary pattern of the computational basis states, using global microwave pulses in combination with an auxiliary rotational state outside the computational basis for site-selective shelving of molecules.

This work is organized as follows: in Section \ref{sec:DescribeHFI} we detail the implementation of Cs imaging at the high magnetic field conditions in which molecules are prepared. In Sections \ref{sec:MoleculeRearrangement} and \ref{sec:SequentialImaging} we demonstrate molecule rearrangement and sequential imaging of multiple rotational states. In Section \ref{sec:SiteSelective} we demonstrate site-selective rotational excitations. 
We conclude with suggested applications and improvements to the fidelities of these schemes in \ref{sec: conclusion and outlook}.

\section{high-field atom imaging}\label{sec:DescribeHFI}

\begin{figure}[ht]
\centering
\includegraphics[width = \columnwidth]{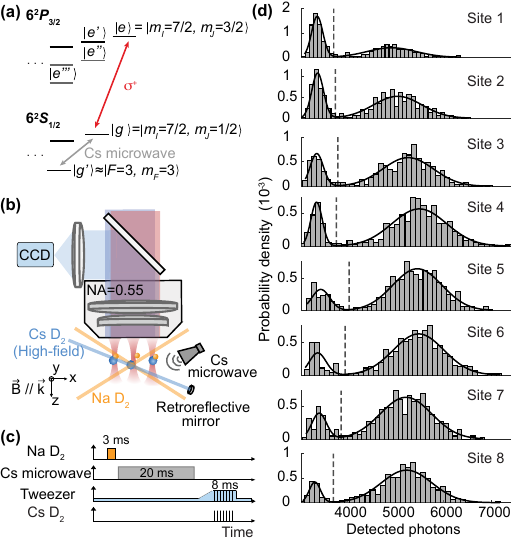}
\caption{
State-selective detection of Cs at 864 Gauss. (a) Simplified energy structure of Cs showing the hyperfine state $\ket{g'}$ that Cs atoms are initialized in and the optical cycling transition between $\ket{g}$ and $\ket{e}$ used for fluorescence imaging. (b) Schematic of the apparatus depicting the beams used for the ejection of sodium atoms and for driving the Cs optical cycling transition. The Cs hyperfine transition is driven by a microwave horn antenna. Fluorescence is collected by a 0.55 numerical aperture (NA) objective and CCD camera. (c) Timing diagram for the imaging process, including Na ejection via near-resonant D2 light, Cs microwave transfer from $\ket{g'}$ to $\ket{g}$, and optical cycling. The tweezer and imaging light are pulsed out of phase since $\ket{e}$ is anti-trapped. (d) Histograms of the collected fluorescence from Cs atoms.}
\label{fig:CsHighFieldImaging}
\end{figure}

Both the rearrangement of molecules and their sequential state-selective readout rely on the detection of atoms at or near the magnetic field at which molecules are assembled. Our preparation of NaCs in its rovibrational and electronic ground state has been detailed previously \cite{Cairncross2021, zhang_optical_2022, picard_high_2023}. In summary, we prepare parallel 1-by-8 arrays of hyperfine ground state Na and Cs atoms near the three dimensional motional ground state of their respective 616 nm and 1064 nm optical tweezers. Na atoms are then adiabatically transferred into the 1064 nm array, such that each 1064 nm tweezer contains, ideally, individual Na and Cs atoms in their motional and hyperfine ground states. Atom pairs are magnetoassociated into a weakly bound molecule via a Feshbach resonance at 864 G and transferred to the rovibrational and electronic ground state via detuned Raman process. 

To detect the unassociated atoms, we extend the low-field state-selective fluorescence imaging of Cs atoms in optical tweezers \cite{Nikolov_Pritchard_2023_StateSelectiveDetection, Chow_Jau_2022_StateDetectionCs} to a high magnetic field, similar to the high-field absorption detection of a cloud of potassium atoms \cite{Ospelkaus2006c}. In principle one could also image Na atoms on the equivalent stretched D2 transition. We opt to image Cs atoms due to the five times stronger polarizability of Cs in the 1064 nm tweezers, which reduces the tweezer intensity required to trap atoms that are heated during imaging. The magnetic field decouples the total angular momentum $F$ of the excited fine structure levels, such that the quantum numbers $\ket{m_I, m_J}$ best describe the excited hyperfine states, where $I$ is the nuclear spin ($I_{\text{Cs}} = 7/2$), $J$ is the total electronic spin, and $m$ denotes their projections along the magnetic field axis. Detection is performed by exciting from the lowest to highest energy ground hyperfine state and optically cycling on the stretched D2 transition $\ket{g} \equiv 6^2\textrm{S}_{1/2}\ket{F=4, m_F = 4}$ to $\ket{e} \equiv 6^2\textrm{P}_{3/2}\ket{m_I = 7/2, m_J = 3/2}$. The optical cycling transition is depicted in Fig. \ref{fig:CsHighFieldImaging}(a), along with nearby states which can have non-zero coupling to non-$\sigma^+$ components of the imaging light. An advantage of the high-field regime is significant detuning protection against populating these off-resonant states, whose composition and energy are provided in Appendix A.

Following molecule creation or dissociation, Cs atoms are initialized in $\ket{g'}_{\text{Cs}}\approx\ket{F=3,\,m_F=3}$ and may share their optical tweezer with Na atoms in $\ket{g'}_{\text{Na}}\approx\ket{F=1,\,m_F=1}$. We must eject Na atoms before exciting Cs to $\ket{g}_\text{Cs}$ to avoid spin-changing inelastic collisions that occur when the atoms are not both in the ground hyperfine state or in spin-stretched hyperfine states \cite{Zhang_Ni_2020_MagnetoassociationTweezer}. As shown in Fig. \ref{fig:CsHighFieldImaging}(b), which depicts the geometry of the apparatus, the Na atoms are heated out of the trap by beams at the D2 transition frequency that are originally used to produce the Na magneto-optical trap (MOT). The Cs atoms are then excited from $\ket{g'}$ to $\ket{g}$ using a microwave horn with negligible off-resonant coupling due to a $h\times 250$ MHz detuning with the next nearest transition. Lastly, we optically cycle on the $\ket{g}$ to $\ket{e}$ transition using a small retro-reflected imaging beam with a $1/e^2$ waist of $\sim$1 mm, collecting fluorescence photons through a 0.55 NA objective onto a charge-coupled device
(CCD) camera. The use of a  small imaging beam helps suppress background noise due to light scattered from the surface of the glass cell into the objective. Vertical polarization is chosen to mitigate the $\pi$ polarized $\ket{g}$ to $\ket{e'}\approx6^2\textrm{P}_{3/2}\ket{m_I=5/2, m_J=3/2}$ transition, which is closest detuned by $h\times67$ MHz. Crucially, none of the 589 nm Na D2 light, the 852 nm Cs D2 light, or the 11.34 GHz microwave measurably affect the NaCs molecules.

The timing diagram of high-field imaging is shown in Fig. \ref{fig:CsHighFieldImaging}(c). We observe complete ejection of Na from the trap within 3 ms. For the microwave transition, instability in the magnetic field on the 10 mG level limits the transfer efficiency given the available microwave Rabi frequency. To maximize efficiency in the presence of this noise, we use an adiabatic rapid passage pulse in which the frequency is linearly swept across a 400 kHz range about the resonance in 20 ms. As the imaging geometry provides little cooling during imaging except for red-detuned Doppler cooling along the imaging beam axis, we increase the peak trap intensity to 1.25 $\text{MW/cm}^2$ to collect sufficient fluoresced photons before atom loss to resolve the presence of an atom over background noise. The optical cycling transition is driven near-resonantly to minimize the time that the molecules, whose loss rate from $\ket{0}$ due to scattering of the trap light is $7~ \text{Hz/(MW/cm}^2)$, need to be held at high depth. Due to scattering, we observe a  5.6(8)\% loss of molecules during high-field Cs imaging. We also strobe the imaging and tweezer light out of phase with a 10\% and 70\% duty cycle, respectively, at 500 kHz to prevent heating due to $\ket{e}$ being anti-trapped in the 1064 nm tweezers \cite{Hutzler2017,Martinez-Dorantes_Meschede_2018_DipoleForceHeating}.

The imaging fidelity is calculated from histograms like those shown in Fig. \ref{fig:CsHighFieldImaging}(d) using a standard method that is described in Appendix B.
In order to analyze the imaging fidelity, we prepare Cs atoms in $\ket{g'}$ co-trapped with Na atoms. The fidelity of sodium ejection and subsequent Cs imaging is verified using the sequence shown in Fig. \ref{fig:CsHighFieldImaging}(c). By averaging the collected signal of the CCD camera for each site over many experimental cycles, we observe two Gaussian distributions: a `dark' peak corresponding to electronic read noise of the camera, and a `bright' peak corresponding to atom fluorescence. The `dark' peak remains unchanged in the absence of imaging light, which indicates negligible scattering of the imaging beam off the glass cell into the CCD camera. We report a $0.38(9)$\% false positive rate and $0.089(2)\%$ false negative rate for Cs atoms prepared in $\ket{g}$, corresponding to an overall imaging fidelity of $99.51(9)\%$. Due to imperfect microwave transfer and state-preparation fidelity, the false negative rate increases to 3.6(7)\% for Cs atoms initialized in $\ket{g'}$, which is the case for molecule rearrangement and sequential imaging.

\section{Molecule Rearrangement}\label{sec:MoleculeRearrangement}
\begin{figure}[ht]
\centering
\includegraphics[width = \columnwidth]{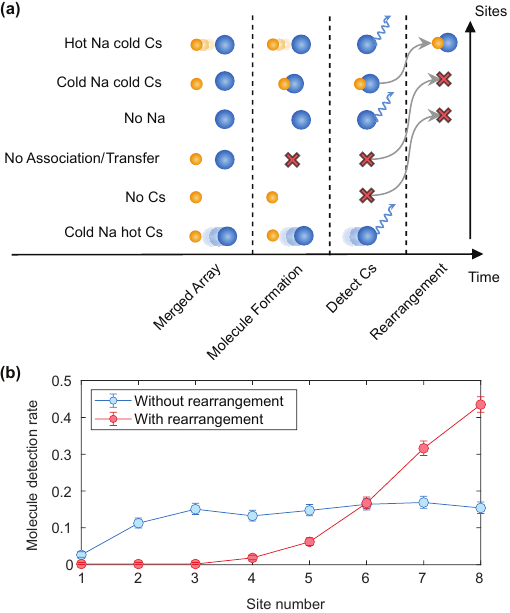}
\caption{Rearrangement of molecules based on the detection of Cs atoms at 864 G. (a) Behavior of the rearrangement algorithm for possible occupations of the 1064 nm tweezers for sites where Na and Cs were both detected before Raman sideband cooling. Sites where Cs is not detected are rearranged. Such sites may not contain a ground state molecule if the Cs atom is lost when the Na and Cs arrays are merged, or if the molecule does not survive Raman transfer to the ground state. (b) Ground state molecule population with and without molecular rearrangement that is detected by Cs imaging at high-field. 
}
\label{fig:FIG3}
\end{figure}
We use the high-field detection of Cs atoms to produce a dense array of molecules without increasing their motional energy. To infer which traps contain molecules we use a combination of three atom images. After the loading and rearrangement of atomic arrays, we detect the initial Na and Cs occupations with fluorescence imaging. The atoms are then Raman sideband cooled to near their motional ground state and the arrays are adiabatically combined. As stated in \ref{sec:DescribeHFI}, Cs atoms are detected again at high-field following molecule production using previously determined thresholds. Molecules may only be found in sites where both atoms were detected at low-field in the first two images, because two atoms are needed to form a diatomic molecule, and where Cs was not detected at high-field in the third image, because the molecules are dark to high-field Cs detection. Such cases correspond to 27\% of the sites that will be preserved by the rearrangement algorithm. 

Three relevant mechanisms can result in the preserved sites not containing ground-state molecules, which are depicted in Fig. \ref{fig:FIG3}(a). First is when atom pairs that associate into Feshbach molecules are lost during detuned Raman transfer, which has a technically limited one-way transfer efficiency of 76(3)\%. Second is that cesium atoms may be lost between low-field and high-field detection, which falsely appears as molecule production. Cs can be lost during the merging of the two arrays (0.7(9)\%), or in spin-changing inelastic collisions with sodium in several ms after merge (1(1)\%) if either atom is prepared in the wrong hyperfine state. A third concern is that Cs atoms may not be detected due to imperfect microwave transfer and state preparation, as described in Section \ref{sec:DescribeHFI}. We estimate based on these mechanisms that 71(3)\% of the preserved sites, or 19(2)\% of all sites, contain a ground state molecule.

The rearrangement of molecules proceeds as follows. Optical tweezers for sites that are not preserved are turned off. Starting sequentially from the rightmost preserved site, tweezers are translated to the rightmost unoccupied site by means of a frequency ramp of the RF tones generating each tweezer. The translation is performed using a minimum-jerk trajectory for the start and end of the ramp, linked by a constant velocity region. At the same time, the RF amplitudes are ramped to compensate for changes in diffraction efficiency across the bandwidth of the acousto-optic deflector (AOD) generating the tweezers, keeping the tweezer depth approximately constant during the move. The hardware for rearrangement has been described previously for atom rearrangement in our system \cite{zhang_optical_2022}. Briefly, images from the CCD camera are processed on a control computer to determine which sites to preserve, and the information is passed to an arbitrary waveform generator that controls RF frequencies sent to an AOD for the 1064 nm tweezers to implement the aforementioned rearrangement protocol. 

The measured ground state molecule population is shown in \ref{fig:FIG3}b. We detect 13.2(4)\% filling of ground state molecules without rearrangement. Accounting for detuned Raman loss and the false negative rate of Cs high-field imaging, we estimate a 18.0(9)\% filling of the array, which is consistent with the prior estimate. Molecular rearrangement significantly improves molecular density on the right side of the array. We detect a factor of 2.8(3) improvement in the filling of the rightmost site, and the detection rate of a pair of molecules in the two rightmost sites improves from 4.3(8)\% to 16(2)\%. This represents a fourfold improvement in statistics for observing the dipolar interaction between two particles. 

In order to quantify any molecule heating induced by the rearrangement process, we dissociate the molecules, unmerge the 616 nm and 1064 nm tweezers, lower the magnetic field, and measure the temperature of the resulting Cs atoms in the 1064 nm trap using the Raman sideband thermometry method described in detail previously \cite{zhang_optical_2022,liu_molecular_2019}. The Cs atom temperature provides an upper bound on the molecule. Using this metric, we detect no resolvable decrease in the motional ground state population as a result of the rearrangement. Further details of the thermometry and measured atom temperature are provided in Appendix C.

\section{Multi-State Readout}\label{sec:SequentialImaging}

\begin{figure}[ht]
\centering
\includegraphics[width = \columnwidth]{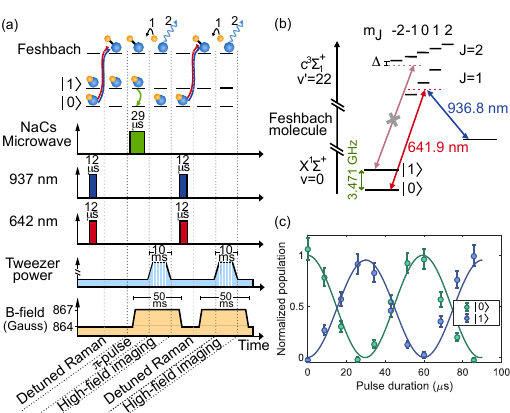}
\caption{Sequential detection of multiple rotational states. (a) Timing diagram showing the sequential dissociation of $\ket{0}$ and $\ket{1}$ and imaging of the resulting Cs atoms.
(b) Energy level diagram of NaCs showing the ground rotational levels, the excited intermediate states used for detuned Raman transfer, and the weakly bound Feshbach state. Population in $N=1$ is protected during detuned Raman transfer by careful choice of single photon detuning. (c) Multi-state detection of rotational population during driven Rabi oscillations between $\ket{0}$ and $\ket{1}$.}
\label{fig:FIG4}
\end{figure}

The ability to perform sequential readout of multiple states of a particle is crucial for qudit-based computation or multi-level quantum simulation applications, due to the quadratic scaling of the number of measurements required for full quantum state tomography \cite{Ma_LaFlamme_2016_QST}. Multi-state readout via fluorescence imaging has  been demonstrated in trapped ion qudits \cite{ringbauer_universal_2022,Low_Senko_2023_IonQudit}, and would be highly applicable to molecules due to their abundance of long-lived rotational and hyperfine states \cite{bao_dipolar_2023}.

In prior work we measured $\ket{0} \equiv \ket{N=0,\,m_N=0}$ population by reversing the adiabatic steps used for molecule formation: two-photon transfer to a weakly bound Feshbach state, magnetodissociation, returning of Na atoms into their original tweezer array, and detection of Na and Cs with fluorescence imaging at 5.5 G. Since neither the Na nor Cs D2 imaging light induces significant scattering of the molecules, we can instead perform multiple rounds of state-selective molecular dissociation and imaging to detect population in $\ket{0}$ and $\ket{1}\equiv\frac{1}{\sqrt{2}}\left(\ket{N=1,\,m_N=-1}-\ket{N=1,\,m_N=1}\right)$, where the rotational ($N$) quantization axis is defined by the magnetic field. Imaging atoms at high-field is advantageous for this application, as it avoids the 150 ms required to ramp the magnetic field down and up between 5.5 and 865 G.

The sequence of field ramps and pulses required for sequential readout of the rotational states $\ket{0}$ and $\ket{1}$ is shown in Fig. \ref{fig:FIG4}(a). The total time needed to detect each rotational state is 50 ms. Population in $\ket{0}$ is converted to a weakly bound molecule via detuned Raman transfer. Prior to its detection, a microwave $\pi$-pulse transfers population in $\ket{1}$ to $\ket{0}$. The weakly bound molecule is dissociated and the resulting Cs atoms, corresponding to the initial $\ket{0}$ population, are detected. Following imaging, a high intensity pulse resonant with the cycling transition is used to clean out any remaining Cs atoms before the next imaging step. Subsequently, the remaining population is dissociated and Cs atoms originally from $\ket{1}$ are also detected. 

The frequency of the 642 nm Raman laser that connects the ground molecular state to the $c^3\Sigma$ potential is carefully selected to prevent population in the $\ket{1}$ state from scattering during the first transfer of $\ket{0}$ to the Feshbach state. The 642 nm laser can couple $\ket{1}$ to the $J=1$ and $J=2$ rotational levels of the excited $v'=22$ manifold, where $J$ denotes the sum of the orbital $N$ and electronic spin $S$ angular momenta of the molecule. We chose the frequency of the 642 nm laser to be $h\times500$ MHz blue detuned from the $\ket{J=1,\, m_J=-1}$, such that direct excitation from $\ket{1}$ would fall between the energy of the $J=1$ and $J=2$ levels, as shown in Fig. \ref{fig:FIG4}(b). The polarization of the two 642 nm and 937 nm Raman beams is therefore set to $\hat{\sigma}^-$ to maximize the Raman Rabi frequency contribution from the closest detuned $\ket{J=1,\, m_J=-1}$ intermediate state. 

To demonstrate sequential imaging, following molecule rearrangement, we apply a resonant microwave pulse and measure Rabi oscillations between $\ket{0}$ and $\ket{1}$, shown in Fig. \ref{fig:FIG4}(c). In the second image we detect 95(3)\% of the population in $\ket{1}$, consistent with expected losses of 5.6(8)\% due to scattering of the tweezer-light during the first imaging step. The multi-state readout procedure provides full-state information on all the molecules in the array, allowing for direct single-shot readout of the diagonal elements of any multi-qubit density matrix.

\section{Site-selective state preparation}\label{sec:SiteSelective}

Site-resolved control of the quantum state of individual particles is necessary for implementing local qubit operations in quantum gates, and to prepare arbitrary initial states in quantum simulators. Towards this goal, local control of fields on the scale of the separation between particles is required. In experiments with neutral atoms in optical tweezers, the conventional approach is to apply a global pulse to drive a transition that is made off-resonant for certain sites using some combination of tightly-focused AC-Stark-shifting beams \cite{labuhn_single-atom_2014,xia_randomized_2015} and manipulation of position-defined sub-ensembles \cite{bluvstein_quantum_2022}.
Recent advances in the trapping of molecules in optical tweezers has extended the technique of local AC Stark shifts to arrays of NaCs \cite{ni_fully_2022} and CaF \cite{bao_dipolar_2023} molecules, a natural choice given the large cm-scale wavelength of microwaves needed for driving rotational transitions as compared to tweezer $\mu$m spacings. 

A drawback of using a computational basis sensitive to differential AC Stark shifts is that intensity noise will result in decoherence. Intensity noise from the optical tweezer is the dominant noise source for molecules due to their generally large tensor polarizability. For this reason, we seek to initialize molecules within a computational subspace that is insensitive to differential AC Stark shifts from the optical tweezer. The ellipticity of the 1064 nm tweezer polarization is tuned to realize a `magic'  state in each rotational level that is free from anistropic polarizability \cite{park_extended_2023}. The lowest two `magic' states, $\ket{0}$ and $\ket{1}$, form our computational basis, whose transition offers a two order-of-magnitude larger spin echo coherence time than the transition between $\ket{0}$ and a non-magic state $\ket{1'}\equiv\frac{1}{\sqrt{2}}\left(\ket{N=1,\,m_N=-1}+\ket{N=1,\,m_N=1}\right)$. Since the computational basis cannot be AC Stark shifted with the 1064 nm tweezer, we make use of the large $h\times5.55(1)\,\text{MHz}/(\text{MW/cm}^2)$ differential AC Stark shift between $\ket{0}$ and $\ket{1'}$.  The $\ket{0}$ population of certain sites is shelved in $\ket{1'}$, allowing the remaining sites to be excited to $\ket{1}$. Shelving avoids the use of an additional near-resonant AC-Stark-shifting beam, which could introduce unwanted scattering that would limit molecule lifetime and state preparation fidelity.

\begin{figure}[ht]
\centering
\includegraphics[width = \columnwidth]{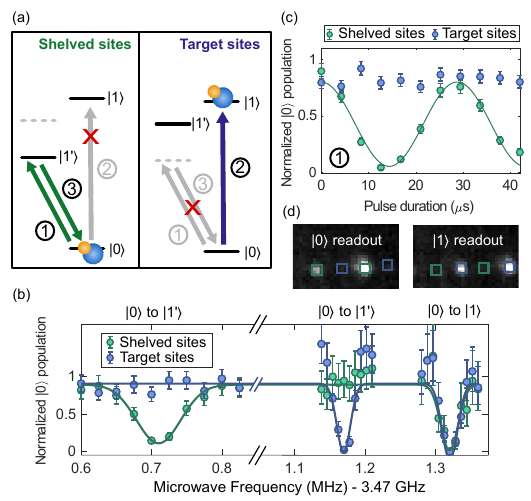}
\caption{Site-selective state preparation in a rotational qubit subspace of $\ket{0}$ and $\ket{1}$. (a) Scheme for an array initialized in $\ket{0}$ showing the shelving of more intense traps to an auxiliary non-magic state $\ket{1'}$ in the $N=1$ manifold. The remaining target sites are driven to $\ket{1}$ with a global microwave pulse. The shelved sites are then returned to $\ket{0}$ with a final resonant pulse. (b) Microwave spectra showing the resolved transition frequencies from $\ket{0}$ to $\ket{1'}$ for the shelved sites (left) and the target sites (center), as well the magic $\ket{0}$ to $\ket{1}$ transition (right) which has the same frequency for all sites. We note that the three peaks are probed with different microwave powers and pulse times. (c) Step 1 of site-selective initialization procedure, showing selective transfer of shelved sites to $\ket{1'}$, while target sites remain in $\ket{0}$. (d) Average sequential readout images corresponding to $\ket{0}$ and $\ket{1}$ population for a site selective initialization of odd and even sites of the array in $\ket{0}$ and $\ket{1}$, respectively.
}
\label{fig:FIG5}
\end{figure}
The scheme for site-selective state preparation is illustrated in Fig. \ref{fig:FIG5}(a). The array is divided into a set of target sites, which will be transferred to the $\ket{1}$ state, and shelved sites, which will remain in $\ket{0}$ at the end of the sequence. For the shelving process, the trap depths of the target sites are reduced to 25\% of the depth of the shelved sites by ramping down the amplitude of the RF tones generating these tweezers. The site-selective initialization is then achieved with a sequence of three microwave pulses:

\begin{enumerate}
    \item With both sites initially in $\ket{0}$, the shelved sites are selectively transferred to the non-magic $\ket{1'}$ state using a resonant microwave pulse. Due to the depth difference between the sites, the corresponding transition for the target sites is off-resonant.
    \item With the shelved sites in $\ket{1'}$, the target sites are then transferred to the magic $\ket{1}$ state.
    \item The shelved sites are then transferred back to $\ket{0}$.
\end{enumerate} 
At the end of the sequence the depths of the target sites are ramped back up. Using this procedure, it is possible to initialize an array of molecules in an arbitrary configuration of states $\ket{0}$ and $\ket{1}$. In \ref{fig:FIG5}(b) we show the resolved microwave transitions to the $\ket{1'}$ state for the shelved and target sites at the depths used for initialization, along with the $\ket{0}$ to $\ket{1}$ which has the same frequency for all sites. The site-selective shelving of some sites in $\ket{1'}$ is shown in Fig. \ref{fig:FIG5}(c). Using this sequence to initialize a four-molecule array into an alternating pattern of $\ket{0}$ and $\ket{1}$, we find we can prepare the target sites in $\ket{1}$ with an average error rate of 4.5(8)\%, and leave shelved sites in $\ket{0}$ with an error rate of 3.7(7)\%. Error rates are determined by comparing the average population of each sub-ensemble measured in the target state and the incorrect computational basis state after the initialization sequence. An average fluorescence image showing the array state after site-selective preparation is shown Fig. \ref{fig:FIG5}(d). In the future, these error rates could be further reduced with the use of composite pass-band pulse sequences, which can reduce the sensitivity of a $\pi$-pulse to small detuning and pulse area errors while suppressing coupling to further off-resonant states \cite{ivanov_composite_2015}.

\section{Conclusion and outlook} \label{sec: conclusion and outlook}

We have demonstrated several SPAM capabilities for molecules associated from constituent atoms in optical tweezers: nondestructive detection of occupation, rearrangement, site-selective initialization, and sequential multi-state detection. Many of these techniques can be generalized to other systems; multi-state  detection is broadly applicable to bulk and lattice platforms, 
and site-selective initialization can be extended to other multi-level systems in optical tweezers. 

Further technical improvements can increase the fidelity of these capabilities. The detuned Raman transfer efficiency---which limits molecule production, the confidence that a rearranged site contains a molecule, and the detection fidelity---can be increased with beam intensity and position stabilization, and more laser power. The false negative rate of high-field Cs detection can be reduced by improving the transfer fidelity between $\ket{g'}$ and $\ket{g}$ with a stronger microwave Rabi frequency. The shelving of molecular population in $\ket{1'}$ for site-resolved initialization can be improved with shaped or dynamical decoupling pulses to overcome noise in the transition frequency and off-resonant coupling to other $N=1$ states. Lastly, molecule loss due to the high tweezer intensity needed for Cs detection may be mitigated by separating and detecting Na atoms in a tweezer whose wavelength differs from 616~nm, which is destructive to ground state molecules due to nearby molecular transitions.

In the near term, the SPAM capabilities presented in this work will  improve our ability to implement and characterize two-particle gates based on the dipole-dipole exchange interaction. Molecule rearrangement has produced a fourfold increase of our pair production rate. Additionally, the $\ket{01}$ two-particle state, initialized site selectively, will interact at twice the rate compared to the state initialized by a global $\pi/2$ pulse \cite{bao_dipolar_2023, Holland_Cheuk_2023_DDI}. Single particle addressability is also a key requirement for realizing a universal gate set with molecules based on the iSWAP entangling gate \cite{Ni2018}. Sequential multi-state imaging will enable faster accumulation of statistics, such as a fourfold reduction in the time required for parity measurements to quantify Bell state entanglement fidelity, or efficient characterization of errors in precision measurement schemes using molecules \cite{fleig_theoretical_2021}. In the long term, advances such as the ones described here will be necessary to use polar molecules as qudit or n-qubit systems for quantum computation and simulation applications.

\section*{Acknowledgement}
 We thank Jessie Zhang for early experimental assistance and discussions. This work is supported by the AFOSR MURI grant \#FA9550-20-1-0323, AFOSR grant \#FA9550-23-1-0538, and NSF grant \#PHY-2110225.\\
 \textit{Note}--- During the completion of this work, we became aware of a concurrent work using RbCs molecules from the Cornish group at Durham University \cite{Rutley_Cornish_2024_SPAM}.

\appendix
\setcounter{figure}{0}
\renewcommand{\thefigure}{A.\arabic{figure}}

\section{Atomic structure}
 The hyperfine structure of the $3^2S_{1/2}$, $3^2P_{3/2}$, $6^2S_{1/2}$, and $6^2P_{3/2}$ fine structure levels of Na and Cs at the conditions used for molecule formation is provided to complement the discussion of high-field imaging in \ref{sec:DescribeHFI}. Atomic parameters are obtained from Ref. \cite{sibalic2017arc}. We assume the atoms experience two external fields: a magnetic field that defines the quantization axis $\vec{B} = 864.1\,\text{G}\, \hat{z}$, and an optical tweezer specified by ellipticity $\chi=0.9553$ and orientation $\psi=0$ with polarization 
\begin{equation}
    \hat{\epsilon}=e^{-i\psi}\frac{\cos(\chi)-\sin(\chi)}{\sqrt{2}}\hat{\sigma}_--e^{i\psi}\frac{\cos(\chi)+\sin(\chi)}{\sqrt{2}}\hat{\sigma}_+ 
\end{equation}
The hyperfine structure is calculated by diagonalizing the bare hyperfine structure, Zeeman shift, and AC Stark shift terms in the Hamiltonian in the $\ket{I,\,J,\,m_I,\,m_J}$ basis and assuming no coupling between fine structure levels. The AC Stark shift is incorporated following the procedure in Ref. \cite{Kien2013Polarizability}. The composition, energy, polarizability, and magnetic dependence of relevant eigenstates are provided in tables \ref{table:CsStates}.I. and \ref{table:NaStates}.II. for Cs and Na, respectively.

\begin{table*}
\label{table:CsStates}
\caption{ Properties of the states relevant to the high-field imaging for Cs, including composition, energy, polarizability, and magnetic field dependence. State composition and energy relative to the fine structure energy are specified at 864.1 G in the absence an external AC electric field. Ground states are best described in the $\ket{F,\,m_F}$ basis, and excited states are best described in the $\ket{m_I,\,m_J}$ basis.}
\begin{tabularx}{0.95\textwidth}{|p{1.5 cm}|X|p{1.5 cm}|p{1.8 cm}|p{1.8 cm}|}
\hline
State label & State composition & Energy [$h\times\text{GHz}$] & Polarizability $\left[\frac{h\times\text{MHz}}{\text{MW}/\text{cm}^2}\right]$ & Zeeman shift $\left[\frac{h\times\text{MHz}}{\text{G}}\right]$\\
\hline\hline
$\ket{g'}$ & $- 0.997 \ket{3,3} + 0.072\ket{4,3}$ & -6.139 & -51.0 & -1.175 \\
\hline
$\ket{g}$ & $\ket{4,4}$ & 5.231 & -59.2 & 1.399\\
\hline
$\ket{g''}$ & $-0.997\ket{4,3} - 0.072\ket{3,3}$ & 4.987 & -58.5 & 1.172\\
\hline
$\ket{g'''}$ & $-0.995\ket{4,2} - 0.099\ket{3,2}$ & 4.731 & -57.7 & 0.926\\
\hline
$\ket{e}$ & $\ket{7/2,3/2}$ & 2.682 & 21.8 & 2.799 \\
\hline
$\ket{e'}$ & $0.998\ket{5/2,3/2}+0.067\ket{7/2,1/2}$ & 2.615 & 21.7 & 2.791\\
\hline
$\ket{e''}$ & $0.998\ket{7/2,1/2}-0.067\ket{5/2,3/2}$ & 0.885 & -9.6 & 0.940\\
\hline
$\ket{e'''}$ & $0.997\ket{7/2,-1/2}-0.075\ket{5/2,1/2}$ & -0.906 & -13.8 & -0.925\\
\hline
\end{tabularx}
\end{table*}
\begin{table*}
\label{table:NaStates}
\caption{Properties of the states relevant to the high-field imaging for Na, including composition, energy, polarizability, and magnetic field dependence. State composition and energy relative to the fine structure energy are specified at 864.1 G in the absence an external AC electric field. All states are best described in the $\ket{m_I,\,m_J}$ basis.}
\begin{tabularx}{0.95\textwidth}{|p{1.5 cm}|X|p{1.5 cm}|p{1.8 cm}|p{1.8 cm}|}
\hline
State label & State composition & Energy [$h\times\text{GHz}$] & Polarizability $\left[\frac{h\times\text{MHz}}{\text{MW}/\text{cm}^2}\right]$ & Zeeman shift $\left[\frac{h\times\text{MHz}}{\text{G}}\right]$\\
\hline\hline
$\ket{g'}$ & $ 0.977 \ket{-1/2,-1/2} - 0.215\ket{3/2,-1/2}$ & -2.046 & -11.0 & -1.273 \\
\hline
$\ket{g}$ & $\ket{3/2,1/2}$ & 1.874 & -11.0 & 1.400\\
\hline
$\ket{g''}$ & $-0.977\ket{3/2,-1/2} - 0.215\ket{-1/2, -1/2}$ & 1.601 & -11.0 & 1.271\\
\hline
$\ket{g'''}$ & $-0.951\ket{1/2,1/2} - 0.310\ket{-3/2,1/2}$ & 1.279 & 11.0 & 1.132\\
\hline
$\ket{e}$ & $\ket{3/2,3/2}$ & 2.461 & 87.6 & 2.799\\
\hline
$\ket{e'}$ & $1.000\ket{1/2,3/2}+0.018\ket{3/2,1/2}$ & 2.434 & 87.6 & 2.800\\
\hline
$\ket{e''}$ & $1.000\ket{3/2,1/2}-0.018\ket{1/2,3/2}$ & 0.818 & 6.7 & 0.933\\
\hline
$\ket{e'''}$ & $1.000\ket{3/2,-1/2}-0.020\ket{1/2,1/2}$ & -0.823 & -37.5 & -0.935\\
\hline
\end{tabularx}
\end{table*}
\section{Quantifying imaging fidelity}\label{sec:imagingFidelityAppendix}
We describe the standard procedure used to determine the fidelity of Cs high-field imaging. Cs atoms are optically cycled on the $\ket{g}$ to $\ket{e}$ transition and fluorescence is detected by a CCD camera via a 0.55 NA objective. Since the signal from each atom is spread across several camera pixels, we sum the photons collected in a 3x3 region of pixels. We observe two distributions in the number of collected counts over many experimental cycles; a lower-count `dark' peak corresponding to background noise, and a higher-count `bright' peak corresponding to the convolution of background noise and atom fluorescence. The background noise is likely read noise of the camera; we observe no change in the background distribution when the imaging beams are turned off, and the `dark' distribution is better modeled by a Gaussian, rather than a Poissonian distribution that is typical for scattered background light.

We proceed to fit two Gaussian distributions to the collected counts and determine a threshold for counts above which we determine a site contains an atom. The threshold is chosen to minimize the fraction of the `dark' Gaussian distribution that is above the threshold and the fraction of the `bright' distribution that is below the threshold. Once an optimal threshold is chosen, these two fractions represent the false positive and false negative rates of detection. Fidelity is then defined as the difference between unity and the sum of these false rates.

\section{Probing molecule temperature}\label{sec:Thermometry}

\begin{figure}[ht]
\label{fig:AppendixRSC}
\centering
\includegraphics[width = \columnwidth]{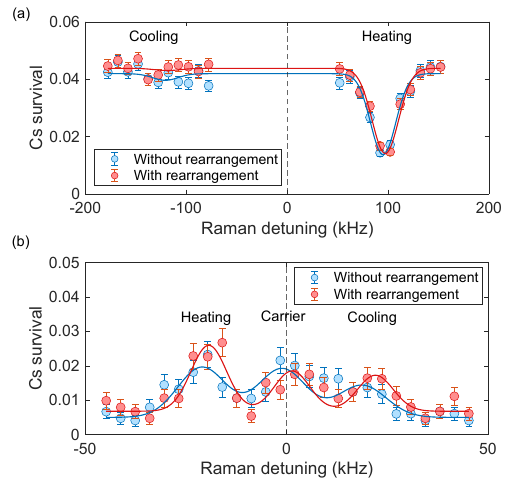}
\caption{(a) Radial sideband thermometry on Cs atoms, starting in the $\ket{F=3,m_F=3}$ state and attempting to drive to heating and cooling sidebands of the transition to $\ket{F=4,m_F=4}$, both with and without the molecule rearrangement step active. (b) Axial sideband thermometry, starting in $\ket{F=4,m_F=4}$ and driving to $\ket{F=3,m_F=3}$.
}
\end{figure}

As described in the main text, we probe the motional temperature of the ground state molecules after rearrangement by first dissociating them and then performing Raman sideband thermometry on the resulting Cs atoms. This method can only provide an upper bound on the temperature of the molecules, as the atoms undergo additional heating following dissociation, most notably during the unmerging of the 1064 nm and 616 nm tweezer arrays. However, atom thermometry can still provide a useful indication of any additional heating induced by the rearrangement process. The thermometry procedure, described in detail previously \cite{Yu2018,liu_molecular_2019,zhang_optical_2022}, involves using a pair of lasers detuned from the Cs D2 transition to drive motional sidebands on the Raman transition between the $\ket{F=3,m_F=3}$ and $\ket{F=4,m_F=4}$ hyperfine states. Atoms in $\ket{F=4,m_F=4}$ are pushed out with resonant light at the end of the sequence, allowing the population in $\ket{F=3,m_F=3}$ to be inferred from the remaining atom survival.

In Fig. A.1., we show Raman sideband spectra for the radial and axial axes of the tweezer trap. The weakly confined axial direction corresponds to the tweezer \textit{k}-vector, and the radial direction is in the plane of the array. To extract the molecule temperature, we fit the heights of the peaks \cite{Yu2018}. In order to constrain the fit, we fix the frequency separation of the peaks based on the measured trap frequencies at the same tweezer depth that we use for Raman sideband cooling. Motion in the radial direction would be particularly deleterious for future work on coherent dipole-dipole interactions, because it corresponds to a linear change in the molecule-molecule separation $r$. Because of the array geometry, $r$ is only quadratically sensitive to axial relative motion. In the radial direction, we measure motional ground state fractions of 92(7)\% without rearrangement and 97(6)\% with rearrangement, which is consistent to within one standard error with there being no radial heating during rearrangement. In the axial direction we measure 39(20)\% ground state population without rearrangement and 54(17)\% with rearrangement, which is also consistent with no heating to within one standard error with no heating. Note that for the axial spectrum we introduce an additional optical pumping pulse to transfer atoms to the $\ket{F=4,m_F=4}$ state before taking the spectrum. At the expense of inducing a small amount of heating, this pulse allows us to more easily resolve the spectrum from the background.

\bibliography{master_ref.bib,master_ref_Dec2023, master_ref_June2023}

\end{document}